\documentclass[12pt,preprint]{aastex}

    \newcommand{\simbad}{{\sc simbad}}
    \newcommand{\ein}{{\it Einstein}\, }
    \newcommand{\einobs}{{\it Einstein Observatory}\, }
    \newcommand{\fuse}{{\it FUSE\, }}
    \newcommand{\fuseobs}{{\it Far Ultraviolet Spectroscopic Explorer (FUSE)\, }}
    \newcommand{\hst}{{\it HST}\ }
    \newcommand{\hstobs}{{\it Hubble Space Telescope (HST)\,}}
    
    \newcommand{\iueobs}{{\it International Ultraviolet Explorer (IUE)\, }}
    
    \newcommand{\ros}{{\it ROSAT\, }}
    
    \newcommand{\kms}{km s$^{-1}$}
    \newcommand{\es}{erg s$^{-1}$}
    \newcommand{\ecs}{erg  s$^{-1}$ cm$^{-2}$}
    \newcommand{\lx}{$L_{\rm X}$}
    \newcommand{\llx}{log $L_{\rm X}$}
    \newcommand{\lbol}{$L_{\rm bol}$}
    \newcommand{\lovi}{$L$(O\,{\sc vi})}
    \newcommand{\ciii}{$L$(977\AA)}
    \newcommand{\lya}{Ly\,$\alpha$}
    \newcommand{\msun}{$M_{\odot}$}
    \newcommand{\teff}{$T_{\rm eff}$\,}
    \newcommand{\acep}{$\alpha$~Cep}
    \newcommand{\iuma}{$\iota$~UMa}
    \newcommand{\dvel}{$\delta$~Vel}
    \newcommand{\omeaur}{$\omega$~Aur}
     \newcommand{\teri}{$\tau^3$~Eri}
    \newcommand{\icen}{$\iota$~Cen}
    \newcommand{\bleo}{$\beta$~Leo}
    \newcommand{\dleo}{$\delta$~Leo}
    \newcommand{\bari}{$\beta$~Ari}
    \newcommand{\roct}{$\rho$~Oct}
    \newcommand{\eeri}{$\epsilon$~Eri}

\shorttitle{{\it FUSE} Observations of A Stars}
\shortauthors{Neff \& Simon}

\begin{document}

\title{\ion{O}{6} Observations of the Onset of Convection Zones
in Main-Sequence A Stars
\footnotemark}

\author{James E. Neff}
\affil{Department of Physics \& Astronomy, College of Charleston, Charleston, SC 29424}
\email{ neffj@cofc.edu}

\and 

\author{Theodore Simon\altaffilmark{\dag}}
\affil{Institute for Astronomy, University of Hawaii, Honolulu, HI 96822}
\altaffiltext{\dag}{Current Address:  Eureka Scientific Inc., 1537 Kalaniwai Place, Honolulu HI 96821}

 \footnotetext[1]{Based on observations made with the NASA-CNES-CSA
    {\it Far Ultraviolet Spectroscopic Explorer (FUSE)}, operated
    for NASA by the Johns Hopkins University under NASA contract
    NAS5--32985.}
    
\begin{abstract}

If magnetic activity in outer stellar atmospheres is due to an interplay 
between rotation and subsurface convection, as is generally presumed, then
one would not expect to observe indicators of activity in stars with $T_{\rm eff}
\gtrsim$ 8300~K.  Any X-ray or ultraviolet line emission from hotter stars
must be due either to a different mechanism or to an unresolved, active, 
binary companion.  Due to their poor spatial resolution,  X-ray instruments 
have been especially susceptible to source confusion.  At wavelengths longward 
of \lya, the near ultraviolet spectra of stars hotter than this putative dividing line are 
dominated by photospheric continuum.  We have used the \fuseobs to obtain 
spectra of the subcoronal \ion{O}{6} emission lines, which lie at a wavelength 
where the photospheric continuum of the mid- and early-A 
stars is relatively weak.  We observed 14 stars spanning a range in $T_{\rm eff}$
from 7720 to 10,000~K.  Eleven of the 14 stars showed \ion{O}{6} emission lines, including
6 of the 8 targets with $T_{\rm eff} > 8300$~K.  At face value, this suggests that activity does 
not fall off with increasing temperature.  However, the emission lines are narrower than
expected from the projected rotational velocities of these rapidly-rotating stars, suggesting 
that the emission could come from unresolved late-type companions.  Furthermore, the 
strength of the \ion{O}{6} emission is consistent with that expected from an unseen active 
K or M dwarf binary companinon, and the high \lx/\lovi\ ratios observed indicate that this 
must be the case.  Our results are therefore consistent with earlier studies that have shown
 a rapid drop-off in activity at the radiative/convective boundary expected at $T_{\rm eff} 
\sim 8300$~K, in agreement with conventional stellar structure models.

\end{abstract}


\section{INTRODUCTION}

The presence of 1--10 million-K coronae and 10--100 thousand-K chromospheres 
among all late-type stars, including the Sun, is generally accepted to be the direct result 
of stellar magnetism.   Spatially-resolved observations of the solar surface, in fact, 
show a nearly linear correlation between X-ray brightness and magnetic flux that
spans almost 12 orders of magnitude (Pevtsov et al. 2003).   It is only in the case 
of the Sun, of course, that magnetic fields have actually been seen in ultraviolet (UV) 
and  X-ray images of its outer atmosphere.  For every other star, magnetic activity can 
at present only be inferred from spectral proxies, e.g., by the detection of UV emission 
lines or coronal X rays.  It is also widely agreed that the origin of such activity among 
the dwarf stars of spectral types late-F through M lies deep within their interiors, through 
a dynamo process in which rotation interacts in an imperfectly understood way with 
turbulent convection.  The process appears to operate most effectively in very young 
stars, which tend to rotate much more rapidly than the Sun, in very low-mass stars with 
very deep convection zones, and in short-period binaries where tidal interactions 
enforce rapid rotation by synchronizing the spin and orbital motions of the stars.    

Whether magnetic dynamos of a similar nature operate in early-type stars and are 
able to power a hot chromosphere and corona in those objects remains an open 
question.  The two most popular mechanisms for heating the outer layers of low-mass 
stars like the Sun---the dissipation of acoustic or magnetoacoustic waves, and microflaring 
of dynamo-generated turbulent magnetic fields---both require well-developed subsurface 
convection zones, which high-mass stars are thought to lack.  According to stellar structure 
models, main-sequence A stars have vigorous convective cores, but the outer envelopes 
of those stars remain entirely radiative except for thin convective layers in the hydrogen 
and helium ionization zones at the base of the photosphere (Christensen-Dalsgaard 2000; 
Browning et al. 2004; Brun et al. 2005).  The computational and theoretical challenge, 
then, is to show that magnetic fields generated by dynamo action within the convective 
core can rise to the surface on a time scale that is short compared with the main-sequence 
lifetime of such a star (MacGregor \& Cassinelli 2003; see, however,  MacDonald \& 
Mullan 2004) and that such a process indeed leads to the formation of a hot chromosphere 
and corona.

The possibility of dynamo activity in intermediate-mass and high-mass  stars has been 
addressed by a variety of observations in recent years.  Observations by the \iueobs and
the \hstobs\,at near-UV wavelengths have detected high-temperature chromospheric 
emission lines as far up the main sequence as $\bv = 0.16$, among the middle-A stars  
(Simon \& Landsman 1997).  Farther up the main sequence, weak emission in the 
near-UV lines becomes much more difficult to detect against an increasingly bright 
stellar continuum.  To extend the search to higher masses it is necessary to 
look to shorter wavelengths where the photosphere of an A star normally appears much 
darker.  Spectra at UV wavelengths shortward of \lya\ have been obtained for a handful 
of middle- and early-A stars by the \fuseobs spacecraft   (Simon et al. 2002).  The 
observations suggest a possible sharp cut-off in chromospheric emission and convection 
zones close to $\bv = 0.12$,  near an effective temperature of $T_{\rm eff}\sim8300$~K.   
Although based on a small sample of stars, 
such a finding is entirely consistent with the predictions of theoretical models for the location 
of  the radiative/convective boundary line in main-sequence stars (Ulmschneider et al. 1996;  
Christensen-Dalsgaard 2000; Kupka \& Montgomery  2002).  

There are two significant weaknesses of the conventional stellar envelope models that are worth 
mentioning here.  First, the models generally assume that the transfer of energy by convection 
in stars with shallow convection zones can be described by standard mixing-length theory;  
and second, although the chemically normal A-type stars may rotate very rapidly (in some 
cases close to their critical velocity), the models ignore the potential side-effects of axial 
rotation. In particular, recent interferometric observations of two rapidly rotating, spin-flattened 
stars, Altair and \acep, provide evidence for a difference of almost 2000~K between the surface 
temperatures at equatorial latitudes and those at the poles  (van Belle et al. 2001, 2006; 
Peterson et al. 2006).  Consequently, the outer envelope of a rapidly rotating star may become 
substantially convective in its cooler equatorial regions while its hotter polar regions remain
completely radiative.  Recent theoretical models by MacGregor et al. (2007) show the same 
sorts of structural effects from rapid rotation, forming a deeper convective envelope with
increasing rotation.  Thus, the notion of a distinct spectral-type or mass-based ``dividing-line'' 
between the main-sequence stars that have radiative envelopes and those that have convective 
envelopes may apply (in a statistical sense) only to the stars that rotate below a certain 
angular-velocity threshold.  

A small minority of normal A stars in the field and in nearby open clusters have also been 
identified as coronal soft X-ray sources.  Cluster A stars have been detected by \ros in deep 
surveys 
of the Pleiades (Stauffer et al. 1994),  the Alpha Persei cluster (Prosser et al. 1996), and the 
Hyades (Stern et al. 1995), in a few instances at luminosities as high as \lx\ $\approx  10^{30}$
\es.  By comparison, the coronae of most (single) late-type stars in the neighborhood of the Sun, 
including the Sun itself,  radiate much less than $\sim10^{29.3}$ \es\ in soft X rays  (Maggio 
et al. 1987;  Schmitt et al. 1995;  H\"{u}nsch et al. 1998).  Because the majority of cluster A 
stars are undetected in X rays, the emission of the detected 
stars could originate not from the A stars themselves, but from hidden late-type binary companions 
or from unrelated neighboring stars that happen to fall  within the X-ray beam.  Long pointed 
observations with \ein and short scans made during the \ros All Sky Survey (RASS) of 
ostensibly single A stars or late-B stars in the field have also yielded a number of strong 
detections (Schmitt et al. 1985; Simon et al. 1995;  Bergh\"{o}fer et al. 1996, 1997;  H\"{u}nsch 
et al. 1998).  Follow-up observations obtained at higher spatial resolution using the HRI 
camera on \ros (Bergh\"{o}fer \& Schmitt 1994) or the ACIS camera on {\it Chandra} (Stelzer 
et al. 2003) have confirmed the A or B star is coincident with the X-ray source in some cases, 
but in other cases have shown that the X~rays come from a nearby star while the A or B  star 
appears totally dark in X rays.  We note, of course, that even when the positions of the X-ray 
source and the A or B star closely agree, the identity of the source may not  be determined 
with complete assurance, since it is normally difficult  (if not impossible) to prove from an X-ray 
image alone that the emission is intrinsic to the A or B star and not from an unknown or 
unresolved (possibly spectroscopic) binary companion.

As one of the central predictions of the stellar structure models, the location along the main 
sequence where convective envelopes give way to radiative envelopes offers an important 
observational test of the physical models of main-sequence stars.  At present, the two most
commonly-used proxies for convection in stars
give conflicting results for the location of that transformation, which need to be reconciled: 
the far-UV observations of chromospheric activity place the transition among the middle-A 
stars, whereas the X-ray observations of coronal emission place it distinctly farther up the 
main sequence, among the very early-A or late-B stars.  It is essential to determine whether 
the radiative/convective  boundary has been correctly identified by the UV observations or 
by the X-ray observations.  However, observations of the chromospheric lines, which require 
spectra from {\it FUSE},  have thus far been limited to a very small sample of A stars.  The 
present study was therefore undertaken to extend the previous \fuse survey, and it  doubles 
the size of the sample over the critical range in \bv\ color and effective  temperature where 
convection, and hence both chromospheric and coronal emission, are expected to vanish. 
Our main observational goal was to measure accurate fluxes for the high-temperature 
subcoronal emission lines  or to set stringent upper limits on the strengths of those lines.

\section{OBSERVATIONS AND EMISSION LINE PROFILE MEASUREMENTS}

\subsection{Target Selection}

Our sample is comprised of 14 main-sequence A stars. All of the stars have  \bv\ 
colors in the range from 0.23 to 0.01 and optical spectra that are indicative of normal 
chemical abundances.  A list of the stars observed with \fuse can be found in 
Table~\ref{tabparam}.  A journal of the \fuse observations is provided in Table~\ref{tabobs}.  
Seven stars from this sample were 
observed in the cycle 1 Guest Observer (G.O.) program of Simon et al. (2002). 
Those data,  denoted in Table~\ref{tabobs}  by the prefix 
A041, have been reprocessed here with updated calibration software (CalFUSE 3.1.3).  
The spectrum of  $\delta$~Vel was obtained from the cycle 5 G.O.  
program of Cheng and Neff (Program ID E075).  The remaining six stars were observed 
by us in cycle 3 (Program ID C038) and are reported here for the first time.  
Following the failure of a reaction 
wheel on {\it FUSE},  which  limited the declination range that could be accessed by the 
spacecraft, we observed two additional A-type stars, $\sigma$ Ara and $\upsilon$ Lup, 
which were chosen to replace stars on our original target list.  However, the bolometric  
luminosities of  both replacement targets were subsequently determined to be much higher 
than those of the other 14 stars in our sample, and consequently they are omitted 
from the discussion that follows.     

We summarize the relevant parameters of all 14 \fuse targets  in Table~\ref{tabparam}.  
The spectral types, photometry, and parallaxes were extracted from the \simbad\ database. The 
projected rotational velocities are from the published literature, in most cases from the 
papers of Abt \& Morrell (1995) and Royer et al. (2002a,b).  The rotational velocity of $\alpha$~Cep 
is from the interferometry analysis of van Belle et al.  (2006),  which yields a larger value 
than those given by Abt \& Morrell (1995) and Royer et al. (2002b), who employed a more 
conventional approach.  The effective temperatures cited for each star were derived from 
four-color photometry using the ``uvbybeta'' procedure from the IDL Astronomy User's 
Library (Landsman 1993),\footnote{http://idlastro.gsfc.nasa.gov/} which follows the precepts 
of Moon \& Dworetsky (1985). Our \teff\ estimates are generally within a few hundred 
kelvins of the values that were derived independently by Allende Prieto \& Lambert 
(1999) from stellar evolutionary calculations and those that were obtained by Sokolov 
(1995) from the slope of the Balmer continuum between 3200~\AA\ and 3600~\AA. 
The dispersion in temperature for individual stars is comparable to the random scatter 
of approximately $\pm250$~K found by  Smalley et al. (2002) in the \teff values that
were determined for a select group of A stars by a variety of techniques, including the 
Infra-Red Flux Method (IRFM), four-color photometry, Balmer spectral line profile fitting, 
and fundamental measurements.  

The X-ray luminosities listed in Table~\ref{tabparam} are from the 
\einobs measurements of Schmitt et al. (1985), the RASS catalog of H\"{u}nsch et al. 
(1998), and the analysis of a number of pointed observations obtained from the \ros 
archives by Simon et al.  (2002).  The \lx\ value for HD\,129791
is based on our reductions and XSPEC\footnote{The XSPEC 
spectral analysis software is available from the High Energy Astrophysics Science Archive 
Research Center  (HEASARC) of the NASA Goddard Space Flight Center  at 
http://heasarc.gsfc.nasa.gov/docs/xanadu/xanadu.html} modeling of a 6.3 ks  {\it Chandra}
observation of that star, which we obtained from the public archives (ObsID 627; J. L. 
Linsky, Principal Investigator).  HD\,129791 is a wide visual binary (CCDM\,14460--4452) 
with a separation of 35\arcsec\ that is easily resolved by the ACIS camera on-board 
{\it Chandra}.  Both the early-type primary star and its much fainter distant companion 
(CD-44\arcdeg9590B) were detected in X-rays (the secondary at $L_{\rm X} = 10^{29.54}$ 
\es).  The other stars were X-ray selected with an 
a priori bias toward high activity levels so as to favor the detection of high-temperature 
emission lines.   They were designated as X-ray sources by H\"unsch et al. (1998), who 
list all of the main-sequence and subgiant stars from the Yale Bright Star Catalog that 
can be identified with likely X-ray counterparts in the RASS survey. The angular offset 
between the optical and X-ray positions for each A star is $\lesssim$10\arcsec, which 
corresponds to less than one-third the instrumental width of the PSPC camera aboard 
{\it ROSAT}.  Thus, the association of the X-ray source with the A star  instead of a 
widely separated companion or a physically unrelated neighbor is very likely,  but 
by no means definitive, as cautionary experience with {\it Chandra} imaging of several 
\ros detected late-B stars has already shown (Stelzer et al. 2003). The RASS X-ray 
luminosities, along with the {\it Chandra} luminosity we 
determined for HD\,129791, \lx~$\gtrsim1\times 10^{30}$ \es, make all of our  \fuse 
target stars much more active in X rays than the Sun (median \lx = 10$^{27.3}$ \es; 
Judge et al.  2003) and, indeed, more active than every star in the young Hyades 
cluster but three (Stern et al. 1995). 

We used \simbad\ as well as the Palomar Sky Survey (POSS) and 2MASS images 
(Skrutskie et al. 2006) to investigate the binary status of each star in 
Table~\ref{tabparam}.  Our working
assumption is that the UV emission of a bright A star should dominate the light in a 
\fuse spectrum and be less subject to the effects of source confusion from a faint binary 
companion than is the detection of X-ray emission from an X-ray image.  The extent to 
which that critical assumption is borne out by the \fuse observations is discussed below.  
Two stars in Table~\ref{tabparam}, HD\,127971 and \bleo, are wide visual binaries, having a faint 
optical companion that is separated by a large enough distance (35\arcsec\ and 40\arcsec, 
respectively) to ensure that the secondary falls outside the \fuse science aperture (LWRS:  
$30\arcsec  \times 30\arcsec$).  The primaries are thus effectively single.  
Three other stars in our sample,  
\omeaur,  \dvel, and \iuma, are noted as close visual binaries in the \simbad\ database.  
In each case, the binary separation is small enough (5\farcs4, 2\farcs6, and 4\farcs5, 
respectively) and the secondary star is bright enough ($m_v =  8.1, 5.1, \mbox{and } 9.5$ 
mags, respectively) that the companion, if it is active, must  be considered a possible 
source of the observed UV and X-ray emission.  
Finally, two stars in our sample are noted as spectroscopic binaries in \simbad.  
\bari\ is a double-lined spectroscopic binary in an eccentric orbit with a 107 day period 
(Tomkin \& Tran 1987). The spectral type and mass of the secondary component are 
estimated by Tomkin \& Tran and also by Pan et al. (1990)  to be late F or early G and 
$\sim1.2$~\msun, respectively.  The difference in brightness between the primary and 
secondary is a factor of 15 or more.  The other spectroscopic binary in our sample is \iuma.  
The visual secondary, a tight pair of 10th magnitude dM stars (the B and C components of 
the optical triple) is the spectroscopic binary in this hierarchical system.  

\subsection{The \fuse Pointings}

The \fuse spacecraft and its instrumentation are described by Moos et al. (2000) and 
Sahnow et al. (2000).  The individual \fuse observations listed in Table~\ref{tabobs} were all
acquired in the normal time-tag mode through the LWRS large science aperture.  The 
raw datasets were processed with version 3.1.3 of  the CalFUSE calibration software 
(Dixon et al. 2007), which included screening the photon lists and adjusting  the calibrated 
spectra to compensate for a variety of instrumental signatures (e.g., to exclude the times of 
burst events).  Two integration times are listed for each star, the first for the full exposure, 
the second for the nighttime portion of the \fuse orbit,  which normally experiences a lower 
level of contamination from terrestrial airglow (Feldman et al. 2001).  For each of the four 
detector segments (LiF\,1A, LiF\,2B, SiC\,1A, and SiC\,2B), we used CalFUSE to extract
spectra from the  ``good time intervals'' of the individual subexposures. 
Before co-adding, we compared the subexposures to ensure that there were no ``drop-outs''
due to light loss out of the aperture.  We also searched for evidence of substantial variation
from one subexposure to the next.  Finally, we constructed light curves from the time-tagged
data to search for flare-like variations of the \ion{O}{6} emission.  We found no evidence
for flaring in any of the exposures.
The resulting spectra were aligned in wavelength, co-added, and, being highly oversampled in 
wavelength, rebinned by 3 pixels to a resolution of $\Delta\lambda =0.039$ \AA\ as 
an aid in measuring the strengths of various emission lines.  A coarser binning by 9 
pixels was also applied to the co-added data in order to construct a broad UV spectral 
energy distribution (SED) for each star. 

\begin{figure}
\epsscale{0.85}
\plotone{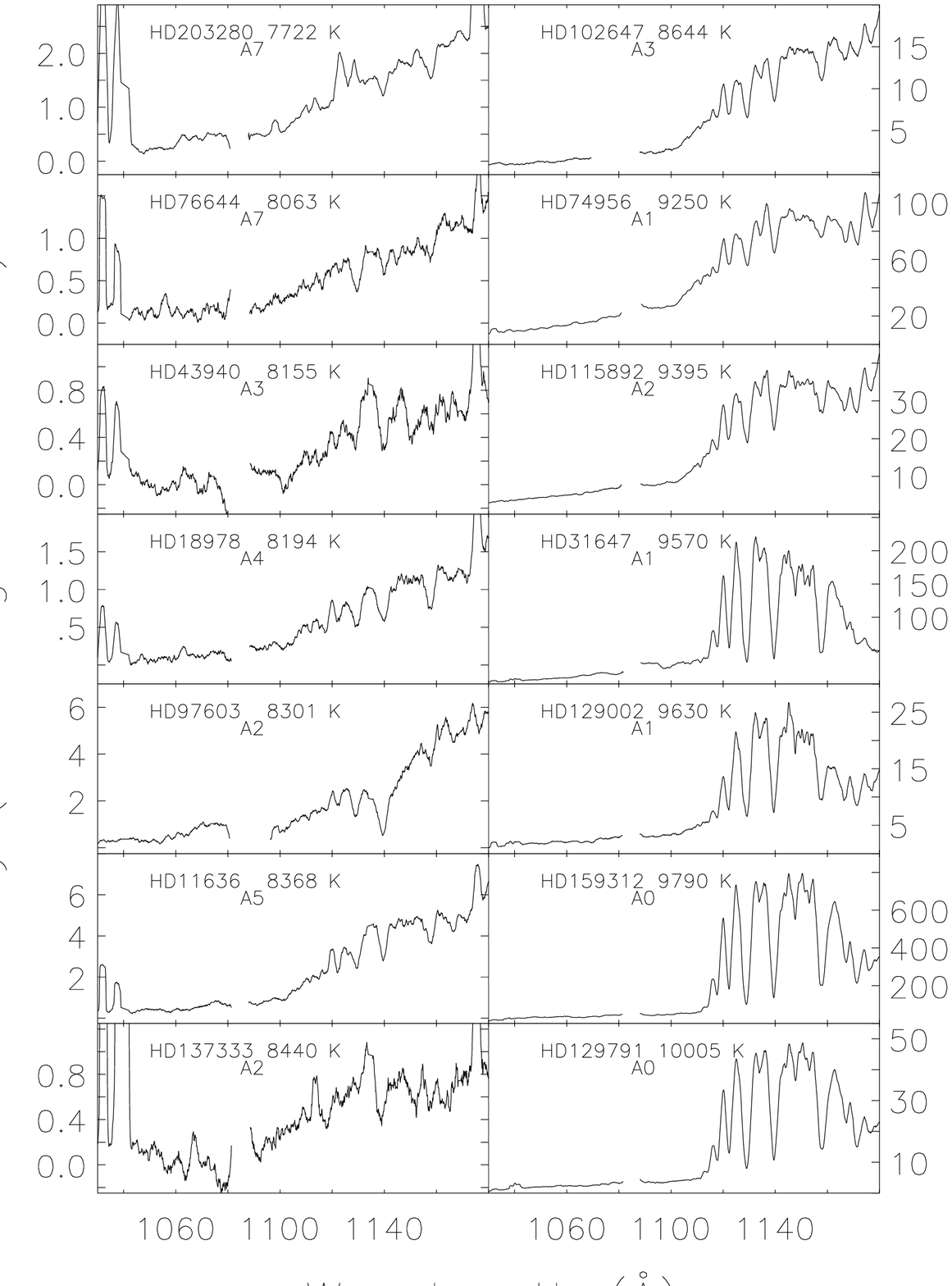}
\caption{ Far-ultraviolet spectral energy distributions of 14 A-type stars.}\label{figsed}
\end{figure}

Figure~\ref{figsed} presents composite spectra for the 14 stars in our full \fuse sample, arranged 
in order of increasing \teff as determined from the four-color photometry.  Shown here
are the LiF\,1A and LiF\,2A detector segments, which have been stitched together to display 
wavelengths from the \ion{O}{6} $\lambda\lambda$ 1032, 1038 doublet to the \ion{C}{3} 
multiplet at 1176 \AA, except for a small gap in spectral coverage at 1080--1090 \AA.\  
Apart from \acep, the short pointings for the remaining 
stars are underexposed in the low-sensitivity SiC channels shortward of the \ion{O}{6} 
lines, so we have truncated the plots to exclude those wavelengths. 
The changes in SED from one star to the next are 
consistent with the ordering in \teff\  established by the optical photometry, despite some 
discrepancies with the \simbad\ spectral types (notably for HD\,43940, which appears 
to have too early a spectral type, and HD\,11636,  or \bari, which may be assigned 
a spectral type too late for its effective temperature).

The ``cleanest'' high-temperature lines in the \fuse spectra of chromospherically active 
late-type stars are normally \ion{C}{3}  $\lambda$977, \ion{O}{6} $\lambda$1032, and 
the \ion{C}{3} $\lambda$1176 multiplet.   \ion{O}{6} $\lambda$1038, the weaker component 
of the \ion{O}{6} doublet, can be blended with both a terrestrial airglow feature and the 
redward component of the stellar \ion{C}{2} $\lambda\lambda$1036, 1037 doublet, as 
can be seen from inspection of the very high quality spectrum of Capella published by 
Young et al. (2001).  Analysis of a moderately high quality spectrum of the A7 V star 
Altair (Redfield et al. 2002) suggests similar line blending issues for that star as well. 
It is quite obvious from Fig.~\ref{figsed} that for most stars the emission in \ion{C}{3} 
$\lambda$1176 is bound to be overwhelmed by bright photospheric continuum.   The 
photospheric continuum is substantially lower at the \ion{O}{6} lines, and the wings of
the \ion{H}{1} Lyman $\beta$ line further enhance the contrast between the emission
lines and the continuum, particularly for the highest temperature targets in our program. 
The usefulness of the short wavelength spectra for the great majority of A-star targets in 
the present study is compromised by low SNR at the \ion{C}{3} 977 \AA\  line. 
Therefore, in the following we focus attention on the \ion{O}{6} features, considering 
principally the stronger $\lambda$1032 component of the atomic doublet.  In solar-type stars,  
these prominent emission features are thought to form in the transition region at a temperature 
of  $\sim$300,000 K.

\subsection{Measurements of the \ion{O}{6} $\lambda\lambda$1032, 1038 Lines}

The integrated \ion{O}{6} line fluxes and their associated 90\% confidence level errors, 
or the appropriate flux upper limits, were measured for each star by modeling the
emission line profile with a single gaussian component and by fitting both linear and 
quadratic terms to the underlying continuum. The best fit was determined by $\chi^2$ 
minimization.  The complete fitting routine was implemented by means of customized 
software, which was written in the Interactive Data Language (IDL), as described by
Neff et al. (1989).  Strong emission lines in the \hst spectra and \fuse spectra of late-type 
stars often show signs of a broad pedestal feature requiring a second gaussian 
component for a proper fit to the observed line profile (Wood et al. 1997; Ake et al. 
2000; Redfield et al. 2002).  However, a double gaussian fit demands a much stronger 
line and a much higher signal-to-noise ratio than was achieved for any of the A stars 
except  \acep, whose spectra (as we are about to show) show no evidence for extended 
line wings.  

\begin{figure}
\epsscale{0.85}
\plotone{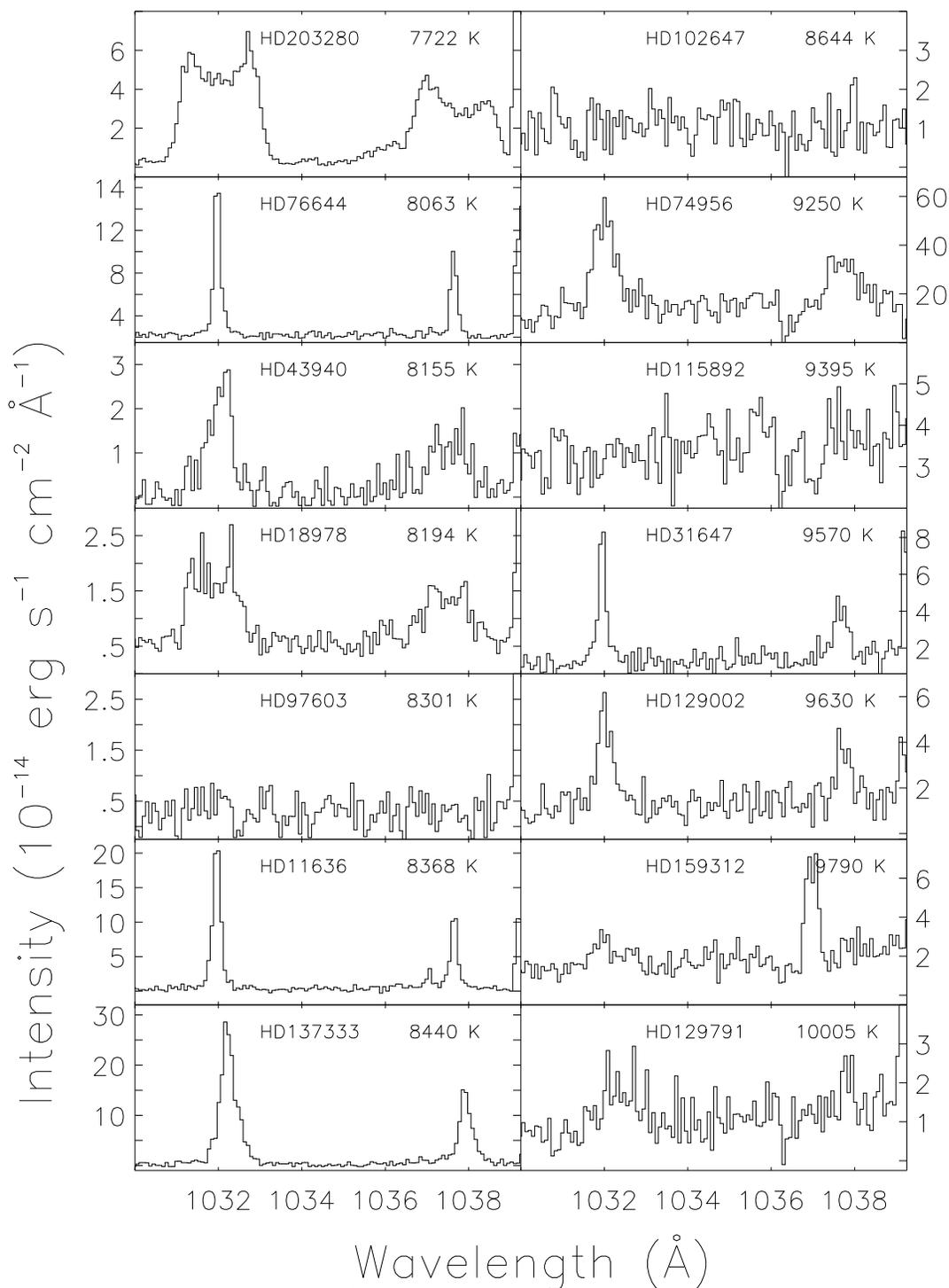}
\caption{\fuse spectra of the \ion{O}{6} $\lambda\lambda$1032, 1038 doublet, rebinned to a 
resolution of 0.04\AA\ per pixel.  \ion{O}{6} emission was detected in 11 of the 14 stars.  
Except for HD203280 and HD18978, the \ion{O}{6} emission lines are narrower than expected
from the star's $v \sin i$.}\label{figo6}
\end{figure}

The results of our measurements are presented in Table~\ref{tabfits}, which
lists the integrated line fluxes for both components of the \ion{O}{6} doublet and
the \ion{C}{3} line, the luminosity of the 
$\lambda$1032 line, the normalized ratio of the line luminosity to the bolometric 
luminosity of the star, and the ratio of the measured HWHM (the half-width at 
half-maximum brightness) to $v \sin i$.  
Detailed spectra of the \ion{O}{6} lines are shown in Figure~\ref{figo6}.  
In all but one case, the integrated fluxes were extracted from the LiF\,1A segment, since 
the Fine Error Sensor camera in the LiF1 channel (FES-A) was the primary camera that was 
most often used for guiding on-orbit prior to July 2005.  Observations made after that date, 
however,  used the alternative FES-B camera in the LiF2 focal plane assembly to maintain 
tracking within the science aperture.  Consequently,  the fluxes of \dvel\ were measured 
instead from the LiF\,2B segment.    
No corrections have been made to the tabulated line fluxes for interstellar extinction, nor 
should they be needed for such nearby stars. 

The stronger $\lambda$1032 component of the doublet is clearly detected in 11 of the 14 stars, 
which range in \teff from 7700 K to as much as 10,000 K. The weaker $\lambda$1038 
component can be measured with confidence in 10 stars.  It is expected to be intrinsically 
half as strong as the short wavelength component of the \ion{O}{6} pair, yet is typically 
somewhat stronger than that, in part because it is difficult to make an accurate measurement 
of weak emission features in such lightly exposed spectra as ours, but also
because of possible contamination from the nearby \ion{C}{2} multiplet.  The 
observed line ratio is particularly discrepant for HD\,159312, for which the red component 
of the doublet is clearly the brighter feature, not only in the LiF\,1A segment but also in 
the LiF\,2B segment.  
The emission is distinctly blueshifted with respect to the nominal wavelength of \ion{O}{6} 
$\lambda$1038 and likely is due to $\lambda$1037 line of \ion{C}{2}.  The asymmetric profile 
of the $\lambda$1038 line of \acep\ (see Fig.~\ref{figo6}) shows unmistakable evidence 
for extra emission from \ion{C}{2} in that star.  
The $\lambda$1037.02 line also appears weakly in emission in the spectrum of \bari.  

For three stars  (\icen, \bleo, and \dleo) there was no measurable \ion{O}{6} line flux.  
Upper limits were determined by fitting a constrained Gaussian with its FWHM set to twice the 
$v \sin i$ of the star, its central peak fixed to match the largest signal in the immediate 
vicinity of the \ion{O}{6} lines, and a background level set at approximately the lower 
envelope of fluctuations in the spectrum at nearby wavelengths.
These three stars with only upper limits 
are among the optically brightest stars in our 
sample and  are by no means the most distant ones observed.  Furthermore, the 13.3 ks 
integration time for \icen\ is one of the longest exposures of all the A stars, second only 
to the 53 ks spent observing \acep.  To within the uncertainties, our independent 
measurements of \ion{O}{6} are in substantial agreement with  those of Simon et al.  
(2002) for the 7 stars that were originally observed by those authors.  

\subsection{Measurements of the \ion{C}{3} 977 \AA\ Lines}

Wherever possible, we measured the \ion{C}{3} 977\AA\  line from the short-wavelength 
SiC portion of the (orbital night only) \fuse spectra (see Table~\ref{tabfits}). 
The peak formation 
temperature of the solar \ion{C}{3} line is $T \sim 80,000$ K, which is appreciably lower 
than that of \ion{O}{6}.  We confirm the previous detections and $\lambda$977 emission line 
fluxes  of Simon et al. (2002)  for \acep\ and  \teri.  However, the flux and emission line 
luminosity of  \ciii $ = 3.0\pm0.2 \times 10^{27}$ \es\  that we obtain for \bari\ are $\sim$40\% 
higher than the earlier published values. 
Our \fuse observations contribute three new detections of \ion{C}{3}:  
HD\,43940, \ciii $ = 9.3\pm2.3 \times 10^{27}$ \es;  
\omeaur, \ciii $ = 22.8\pm5.7 \times 10^{27}$ \es; 
and 33~Boo, \ciii $ = 5.3\pm0.9 \times 10^{27}$ \es.  
The same three stars were also firm detections in \ion{O}{6}.
%

\subsection{Interstellar and Photospheric Absorption Lines}

Although the absolute wavelength scale of \fuse is not well determined, the relative wavelength scale
within a particular detector segment of an observation should be accurate to $\pm5$ \kms\ 
(Redfield et al. 2002).  In order to determine whether the narrow emission lines fall at the wavelength 
expected for the target stars, we searched the LiF1A/2B spectra for interstellar absorption lines or 
photospheric absorption lines that could be used to establish the absolute wavelength scale.  A 
significant offset from the A star's radial velocity would suggest that the emission originates from a 
companion or foreground star.  Because our exposure times were optimized to detect line emission, the 
continuum was always underexposed (see Fig.~\ref{figsed}).  Consequently, in no case  were we able 
to identify suitable photospheric absorption lines in the LiF1A/2B spectra.  

In several cases, we were able to measure the $\lambda$1036.34 \ion{C}{2} interstellar absorption, 
but it provided an extra constraint only for $\delta$~Vel.  For that star, the measured \ion{C}{2} 
wavelength was redshifted by  0.09 \AA\ from the laboratory wavelength, while the measured \ion{O}{6} 
wavelength was redshifted by  0.11 \AA.\  $\delta$~Vel has a measured radial velocity of +2.2~\kms,\ 
and the velocity of the local interstellar cloud in this line of sight is predicted to be +1.8~\kms\ 
(Lallement \& Bertin 1992).   Therefore, the \ion{O}{6} emission of $\delta$~Vel is centered 
within 5 ~\kms\ of the expected wavelength of the A star. 

\section{Results}

\subsection{Line Flux v. Effective Temperature}

\begin{figure}
\epsscale{0.95}
\plotone{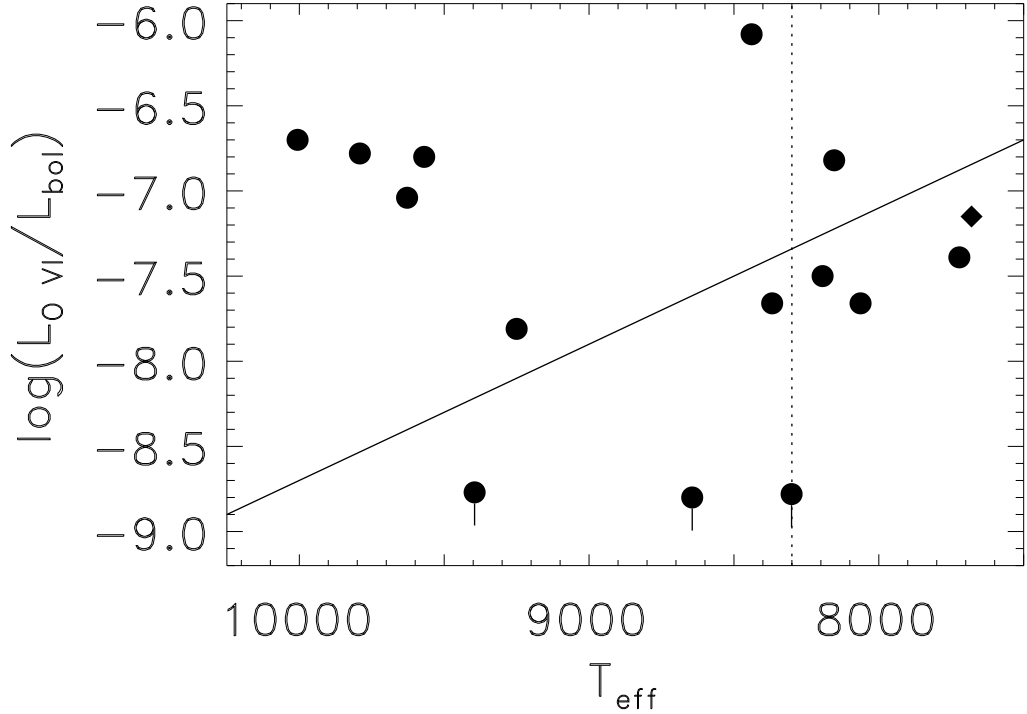}
\caption{Measured \ion{O}{6} emission line luminosity, normalized to the bolometric luminosity, 
plotted as a function of \teff. 
{\it Filled Circles:} A-type stars observed or re-measured in this survey.  Three of these are plotted
as upper limits. The circled points lying below the solid line were originally observed by 
Simon et al. 2002.
{\it Diamond:} the A7 star Altair, data from Redfield et al. 2002.  
The vertical dotted line at $T_{\rm eff}=8300$ is a hypothetical boundary line that separates
stars with sub-photospheric convective layers from those with purely radiative outer envelopes.} \label{figfluxtemp} 
\end{figure}

A plot of the normalized \ion{O}{6} $\lambda$1032 luminosity, ${\cal R} =$ \lovi/\lbol, 
versus $T_{\rm eff}$ is shown  for the entire sample of A stars in Figure~\ref{figfluxtemp}.  The 
points located below the slant line include the seven A-type stars that were originally observed 
by Simon et al. (2002) and also the late-A star Altair, which is denoted by the diamond 
symbol;  the points situated above that line are the X-ray selected target stars that we 
have observed.  In the way of comparison, the normalized $\lambda$1032 luminosity 
of the quiet Sun is ${\cal R}_{\odot}  \approx 10^{-7.1}$  (Simon et al.  2002).  The vertical 
dashed line at $T_{\rm eff}=8300$~K represents the radiative/convective boundary line according 
to the conventional stellar envelope models for main-sequence stars.  

It is obvious that the two sub-samples exhibit a striking difference in behavior.  On the 
one hand, the sample observed by Simon et al. shows an abrupt fall-off in line emission 
precisely at the location of the theoretical boundary line.  Upper limits on the normalized 
luminosities of the stars on the high-temperature side of the boundary are 50 times lower 
than solar, and  20 times lower than the luminosities of A stars on the low-temperature side. 
On the other hand, in the X-ray selected group we have observed,  emission at  the solar 
level is detected up to the earliest spectral type (i.e., A0 V) and the highest \teff (=10,000~K) 
within our sample.  The detection of \ion{O}{6} in the spectrum of HD\,129791, our hottest 
star, is only moderately significant, a slightly less than $4\sigma$ result.  However, as is 
clear from Fig.~\ref{figo6}, there is no question that  \ion{O}{6} emission is present in the spectrum 
of 33~Boo, which is just slightly later in spectral type (A1~V) and just a bit cooler in \teff  
(9630~K) than HD\,129791.  Moreover, as we noted in the previous section,  33~Boo was 
also detected in \ion{C}{3} $\lambda$977 (at  approximately half  the solar normalized 
\ion{C}{3} luminosity), which confirms the apparent activity of this star.  

\subsection{UV Flux v. X-Ray Flux}

\begin{figure}
\epsscale{0.95}
\plotone{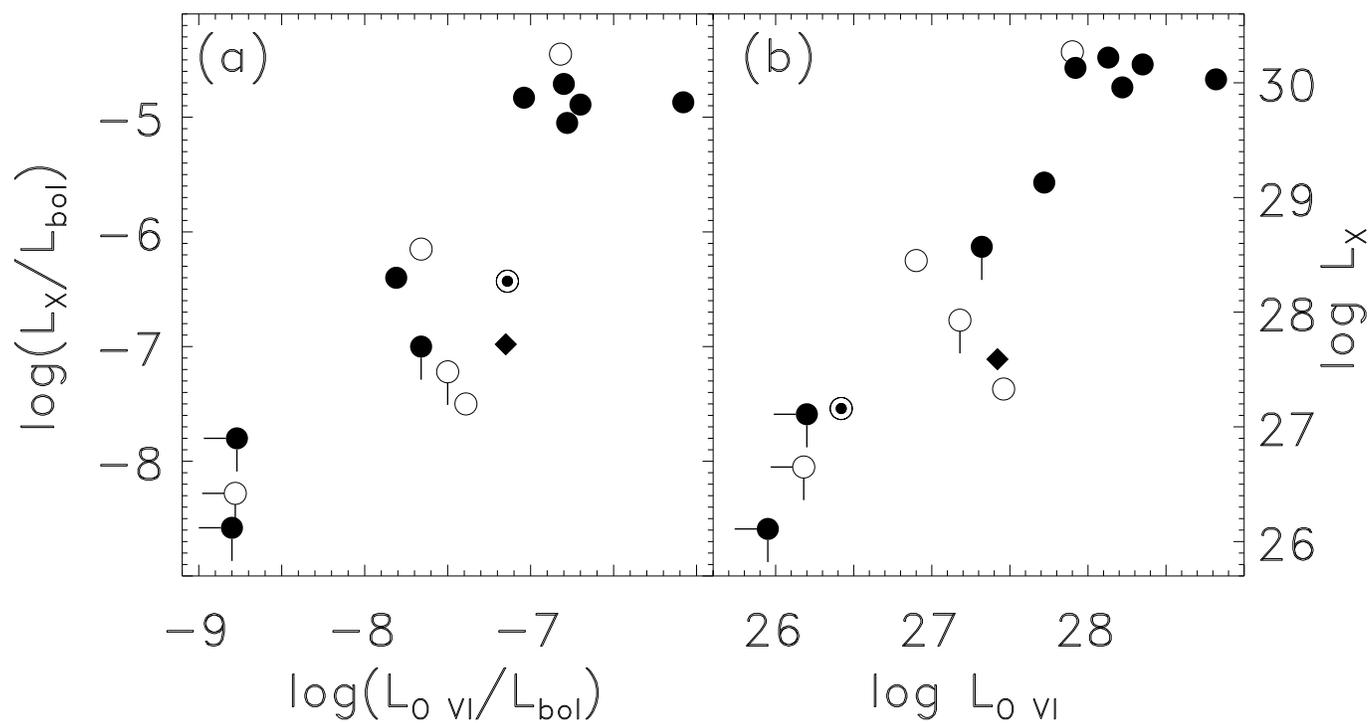}
\caption{(a) Correlation between \lx/\lbol\ and \lovi/\lbol.  
Also plotted are the Sun (circled dot) and Altair (diamond symbol).  The open
circles denote the stars with $T_{\rm eff}\le 8300$, i.e., those to
the right of the dividing line in Fig.~\ref{figfluxtemp}.
(b) A similar correlation is seen between \lx\ and \lovi.}\label{figfluxflux}
\end{figure}

The tight correlation between coronal X-ray emission and the strengths of various 
chromospheric emission lines is well established among stars with spectral types later 
than F5 (e.g., Ayres et al. 1981).  Similarly, for the early A stars in the present study, 
Figure~\ref{figfluxflux} shows a strong relationship between the  \ion{O}{6} line luminosity and X-ray 
luminosity, and an equally pronounced trend in the luminosity normalized to \lbol.  There is no 
significant difference in this relationship between the stars with $T_{\rm eff} \le 8300$~K
(open circles in Fig.~\ref{figfluxflux}) and those with $T_{\rm eff} > 8300$~K 
(filled circles in Fig.~\ref{figfluxflux}).

\subsection{UV Line Widths v. v sin i}

Both Wood et al. (1997) and Redfield et al. (2002) identify the excess broadening of the 
narrow components in the UV emission line profiles of late-type stars with wave 
motions or subsonic turbulence in the transition region, and interpret the highly
supersonic nonthermal broadening of the wide pedestal features in terms of stellar
microflares.  The  role such mechanisms might play in heating the outer atmosphere 
remains unclear.  A comparison of the decomposition of the \ion{Si}{4} $\lambda$1394 
line profile of $\alpha$~Cen\, A and the  \ion{C}{4} $\lambda$1548 profile of AU~Mic 
by Wood et al. (1997) with the analysis of the \ion{C}{3} $\lambda$977 and \ion{O}{6} 
$\lambda$1032 profiles of those same stars by Redfield et al. (2002) demonstrates 
that it is much more difficult to establish the presence of broad line wings in \fuse 
spectra than in \hst spectra.  The SNR of our A star spectra is far from adequate for 
that purpose.

In contrast with the late-type stars, the widths of the \ion{O}{6} lines of the A stars
measured here are generally narrower than the $v \sin i$ values of the stars.
All but two of the observed velocity half widths of \ion{O}{6} are less than 
the $v \sin i$ value (see Table~\ref{tabfits}).  If the chromospheric emission 
covers the star uniformly, we would expect the observed emission line width to be
$v \sin i$ (or larger).  A line of subrotational width could be produced if the 
activity were concentated at high latitudes.  Large polar spots are commonly observed
in Doppler images of magnetically-active stars (e.g. Strassmeier et~al. 2004).  
On the other hand, we would expect the
effective temperature of a rapidly-rotating A star to be
higher at high latitudes (e.g. Peterson et~al. 2006).
If the emission lines arise in the cooler, low-latitude regions, we would expect 
to see rotationally-broadened emission lines.
The narrow emission lines
could also arise from a more slowly-rotating binary companion or foreground
late-type star.  We further discuss this possibility in later sections.

In two cases, \acep\ and \teri, the widths of the observed line profiles are
in excess of the rotational broadening, and the profiles are not well fitted with
a single gaussian emission component.   The fluxes, line widths, and uncertainties 
given in Table~\ref{tabfits} were determined from a series of multiple gaussian fits
and from a numerical integration of the flux above a background fit.
The \ion{O}{6} spectrum of \acep\ shown in Fig.~\ref{figo6} bears a very close resemblance to 
the \fuse spectrum of Altair ($\alpha$~Aql), which Redfield et al. (2002) interpret as 
rotationally broadened and limb brightened.  The physical parameters of the two stars 
are also quite similar. As we noted earlier, the prominent blueward asymmetry in the 
emission profile of  $\lambda$1038 is attributable to an added contribution from 
\ion{C}{2}.  A similar distortion can be seen in the high signal-to-noise spectrum of the 
\ion{O}{6} line of Altair,  which Redfield et al.  decompose into separate 
Gaussian components.  The box-like shape of the profiles of both stars is reminiscent 
of their \ion{Si}{3} $\lambda$1206 line in \hst  spectra (Simon \& 
Landsman 1997).  The profile of the $\lambda$1032 line of \teri, a somewhat slower 
rotator than either Altair or \acep, exhibits a narrower width but a similar flat-top shape,
although clearly at a much lower SNR. 

\subsection{Line Flux v. v sin i}

UV and X-ray emission of a deeply 
convective late-type star increases with its axial rotation rate except at the very highest 
rotation speeds,  while the emission of a thinly convective late-A or early-F star is 
entirely independent of its projected rotation speed (Ayres \& Linsky 1980; Simon 
\& Fekel 1987;  Simon \& Landsman 1991).  This difference in the behavior of stars
earlier and later than spectral type F5 has led to the suggestion that the activity of
these two groups of stars may be produced by different mechanisms (Wolff et al. 
1986).  The X-ray and \ion{O}{6} luminosities or normalized emission luminosities
of the early A stars under investigation here likewise show no dependence on 
$v \sin i$  (cf Tables~\ref{tabparam} and \ref{tabfits}).  Although our sample is relatively small,  
the range in velocity is fairly wide, from a low value of 70 \kms\  
to a top value of 280 \kms.  
      
Consistent with the outcome of earlier studies, we find that both the coronal X-ray luminosity 
and the normalized X-ray luminosity of the A stars in this work are independent of the stellar 
rotation rate, $v \sin i$.  The same appears to be  true of the UV emission in \ion{O}{6}.  
Therefore, if rapid rotation has the effect of deepening the outer convection zone of A stars, 
as suggested by the interferometric observations of Altair and \acep\  (e.g., van Belle et al. 
2001, 2006) and also by the stellar models of MacGregor et al. (2007),  and of promoting 
the formation of a convectively heated chromosphere or corona,  we can point to no empirical 
evidence for that effect in the observations presented here.  At the same time,  if the activity of 
the A stars is powered by a shear dynamo that operates in the largely radiative portions of the 
envelopes of these A stars, as proposed by MacDonald \& Mullan (2004)  for the 
O and B stars, we would again expect to find a trend of increasing X-ray and UV emission with
increasing rotation rate, which is not evident in our results. 

\section{Discussion}

The possibility remains that the X-ray as well as the UV emission we observe from the A stars 
is produced by unrecognized and heretofore undetected companion stars.  One  example of 
such possible source confusion is provided by the active Hyades F star, 71 Tau  (HD\,28052), 
whose chromospherically active companion was discovered in near-UV emission lines only 
by virtue of the high spatial resolution of \hst  (Simon \& Ayres 2000).  Inspection of 
Fig.~\ref{figo6} and Table~\ref{tabfits} suggests that relatively narrow UV emission lines 
predominate over the obviously 
rotationally-broadened profiles of \acep\  and \teri, suggesting that the narrow 
emission features may originate not from the A stars but from more slowly rotating late-type 
companions. 
 Consider, for example, the 9 stars with \teff values hotter than 8300~K.  
\icen\ and \bleo\ lack \ion{O}{6} emission, and the measurement of HD\,159312 is
uncertain because only the $\lambda$1032 line of \ion{O}{6}
was detected (the $\lambda$1038 line was likely obscured by \ion{C}{2} emission).  
Of the remaining 6 stars, only HD\,129791, an extremely rapid 
rotator like \acep, has broad lines, and in that case the evidence is not definitive because 
the line emission is weak and the observation is underexposed.  The remaining five  stars  
(33~Boo, \omeaur, \dvel, \roct, and \bari) all have narrow UV lines, of which three (\omeaur,  
\dvel, and \bari) are already known to have close optical or spectroscopic companions.  

Any hidden low-mass secondary able to produce the powerful coronal emission of the 
X-ray selected A stars we have observed with {\it FUSE} would have to be 
counted among the most active late-type stars in the solar neighborhood.  For simplicity, 
we will assume in the following discussion that any such active secondary is a very cool
dwarf, either a dK or a dM star.  Young, rapidly rotating F and G dwarfs are also known to 
have X-ray luminosities as high as 10$^{30}$ \es, but owing to their higher mass would 
most likely have made their presence known by inducing radial velocity variations in the
spectrum of the primary.  As exemplars of possible active secondaries, we choose AB~Dor, 
a rapidly rotating early K dwarf ($v \sin i = 80$ \kms; Zuckerman et al. 2004), which is the 
namesake of the AB Dor Moving Group (L\'{o}pez-Santiago et al. 2006), and AU Mic, a 
very young, nearby dM1e star with a dusty disk that is a likely member of the $\beta$~Pic 
Moving Group (Barrado y Navascu\'{e}s et al. 1999).   Both stars are luminous X-ray sources
(Garc\'{i}a-Alvarez et al. 2005; H\"{u}nsch et al. 1999), with \lx $\approx 10^{30}$ \es, and
also strong sources at ultraviolet wavelengths (Wood et al. 1997; Ake et al. 2000;  Redfield 
et al. 2002).   The \ion{O}{6} fluxes and emission line luminosities of both stars are listed in 
Table~\ref{tabfits} (original data from Redfield et al. 2002).  

The UV emission line luminosities of AB~Dor and AU~Mic
are a suitable match to the \lovi\ values of the active A stars that we have observed with \fuse
in all but 3 cases. The \lovi\ values for HD\,129791 and \roct\ are unquestionably on the high side, 
but the SNR in the spectrum of HD\,129791 is quite modest and the detection of its  \ion{O}{6} 
emission is far from the most secure one in our survey.  
%
%
The \ion{O}{6} luminosity of \iuma, a known binary and a star that is on
the convective side of the radiative/convective boundary, is considerably weaker than that of 
either AB~Dor or AU~Mic.  Instead, it is much more similar to the luminosity of the moderately 
active K2~V star \eeri,  to which Redfield et al. (2002) assigned a flux that is equivalent  to \lovi\ 
$ \approx 5\times 10^{26}$ \es.  The X-ray luminosities of the two stars are also very similar, 
\lx = $10^{28.4}$  \es\ for \iuma\  and  \lx = $10^{28.3}$ \es\ for  \eeri\  (H\"{u}nsch et al. 1998).   
There is currently no reason to believe that a star as cool as the A star primary in \iuma\  cannot 
support a chromosphere as well as a corona, but if the high-energy emission observed in this 
system does come from a companion star, then the secondary (or the secondaries) needs to be 
mildly but not excessively active.    


The foregoing results are summarized in Table~\ref{tabtruth} in the form of a ``truth table.''  A 
check mark 
next to the name of a star signifies that the proposition stated at the head of a column is TRUE, 
otherwise it is FALSE or not determinable from the available data.  If we exclude both \acep\ and 
\teri, for which there is little doubt that the X-ray and UV activity is intrinsic to the A star, the 
detection of strong X-ray emission that is consistent with an active dK/dM star then appears to 
be a sufficient condition for the presence of UV emission at an intensity level that is also expected 
for a dK/dM star; moreover, with but one exception (i.e., \bari, which is a known binary), it is also 
a necessary condition. 

The commonality of the narrow UV line widths is suggestive of a binary origin of the observed
emission, but is not conclusive.  Narrow lines could also arise from a high-latitude distribution
of active regions, perhaps related to polar spots often seen in the photospheres of active,
late-type stars.  
A further clue as to the origin of the emission can be found,
however, in a comparison of the X-ray luminosity with the \ion{O}{6} luminosity.  The ratio of the 
two, \lx/\lovi, serves as a rough gauge of the differential emission measure between the middle 
chromosphere and the corona, and is known to be much larger for an extremely active star like 
AB~Dor or AU~Mic than for a low-activity star like the Sun.   For the former the ratio is $\sim$130 
and $\sim$225, respectively; for the Quiet Sun it is $\sim$5; and for a late A-type star like Altair, 
which is an example of a higher-than-solar mass star with an ``X-ray deficit'' (Simon \& Drake 
1989), it is $\sim$1.  Among the 9 hottest stars in our sample of A stars whose effective 
temperatures formally place them above the radiative/convective dividing line at 8300~K, six 
have supersolar  \lx/\lovi\ ratios in the range of 25--160, only one of them (the double-line 
spectroscopic binary \bari) has a solar or subsolar ratio, and two (\icen\ and \bleo) have only 
upper limits in both values and hence an indeterminate  X-ray-to-UV luminosity ratio.  Among 
the 5 stars with \teff $<8300$~K, two have supersolar ratios (HD\,43940, 235; \iuma, 30), two 
others have solar or subsolar ones (\teri\ and \acep), and one has dual upper limits and thus 
an undefined ratio (\dleo).

\section{Conclusions}

Of our 14 main-sequence A-type stars, 11 exhibit detectable \ion{O}{6} emission, while  
restrictive upper limits were determined for the other 3 stars.  The stars with 
\ion{O}{6} span the entire range of $T_{\rm eff}$, including the range 
above the presumed convective/radiative ``dividing line'' around 8300~K.  
If this sample is representative, and if the emission indeed arises from the A star, 
then current models of stellar activity must be revised to explain magnetic activity 
in stars without a substantial convective zone.  However, we present several lines of evidence 
that lead to the conclusion that the emission from the higher-temperature stars in 
our sample is more likely due to very active late-type dwarf binary companions.   
The results of our expanded sample are therefore consistent with the observational 
findings of Simon et~al. (2002), which demonstrate a pronounced decrease in UV 
emission at $T_{\rm eff} > 8300$~K, and also with the theoretical predictions of 
the current standard models of stellar structure.  



\acknowledgments

This research has made use of the \simbad\ database, operated at {\it CDS}, Strasbourg, 
France, and is based in part on observations from the public archives of the {\it Chandra
X-Ray Observatory}.  
J.N. acknowledges support by NASA grant NNG04GK80G through 
the \fuse guest observer program to the College of Charleston.  
T.S. acknowledges support by NASA grant NAG5-12198 through 
the \fuse guest observer program to the University of Hawaii.  



{\it Facilities:} \facility{FUSE}, \facility{CXO (ACIS)}.


\clearpage

\clearpage

\clearpage

\clearpage

\clearpage
\begin{deluxetable}{lrlrccrrccrrcr}
\tabletypesize{\scriptsize}
\rotate
\tablecaption{Properties of the Stars Observed With {\it FUSE}\label{tabparam}}
\tablewidth{0pt}
\tablecolumns{14}
\tablehead{
\colhead{ Name }&  \colhead{ ~HD } & \colhead{ Sp. Ty.~ ~} & \colhead{ $\upsilon\,\sin i$ }  &
\colhead{ $V$ }   &  \colhead{ \bv }   &  \colhead{ $\pi$ }         &  \colhead{ $M_v$ }               & \colhead{ $L_{\rm bol}/L_{\odot}$ } &
\colhead{ $\beta$ }  & \colhead{ $b-y$ } & \colhead{ $T_{\rm eff}$ } &  \colhead{\llx} & References  \\[-3pt]
     &      &     &  \colhead{(\kms)}  &   &     &  \colhead{(mas)}  &    &    &     &     &  \colhead{(K)}  & \colhead {(\es)}  &  }
\startdata
HIP 72192\tablenotemark{a} &  129791 &  ~A0 V  &  280\phm{XX}  & 6.91  & 0.05 &  7.72  &  1.35  &  28.9  &  2.867  &  0.035  &  10,000 & 30.16 & 4, 5, 10  \\
HR 6539  &  159312  &  ~A0 V  &  \nodata\phm{W} & 6.48  & 0.01 &  9.64  &  1.40  &  26.4  &  2.886  &  $-$0.005  & 9790 & 29.96 &  2 \\
33 Boo  &  129002  &  ~A1 V  &  95\phm{XX}  & 5.40  &  0.03 &  16.56  &  1.49  &  23.6  &  2.910  &  0.000  &  9630  & 30.13 &  1, 2, 5  \\
$\omega$ Aur\tablenotemark{b} & 31647 &  ~A1 V  &  110\phm{XX} &  4.99 & 0.02  &  20.50  &  1.55 &  22.2  &  2.900  &  0.006  &  9570 & 30.22 & 1, 5  \\
$\iota$ Cen  & 115892 & ~A2 V  & 75\phm{XX}  &  2.70 & 0.09 & 55.64 & 1.43 & 24.1 & 2.901 & 0.004 & 9395  & $<$27.11\phm{X} & 3, 8  \\
$\delta$ Vel\tablenotemark{b,\,{\rm  d}} &  74956 & ~A1 V  &  150\phm{XX} & 1.95 & 0.05 &  40.90 & 0.01  &  87.1  &  2.876  &  0.034  &  9250 & 29.13 & 2, 4, 5  \\
$\beta$ Leo\tablenotemark{a}  & 102647 & ~A3 V  &  115\phm{XX}  & 2.14 & 0.09 & 90.16 & 1.92 & 14.0 & 2.899 & 0.043 & 8645 & $<$26.11\phm{X} & 1, 5, 6  \\
$\rho$ Oct  & 137333 &  ~A2 V   & 150\phm{XX} &  5.58 &  0.11  &  15.02 & 1.46 & 20.9 & 2.887 & 0.072 & 8440 & 30.03 &  2, 3  \\
$\beta$ Ari\tablenotemark{c}  & 11636  & ~A5 V  &  70\phm{XX} & 2.64 & 0.13 & 54.74 & 1.33 & 23.3 & 2.879 & 0.059 & 8370   & $<$28.57\phm{X} & 1, 5, 7  \\
$\delta$ Leo & 97603  & ~A4 V  & 180\phm{XX}  & 2.56 & 0.12 & 56.52 & 1.30 & 23.9 & 2.869 & 0.067 & 8300 & $<$26.65\phm{X} & 1, 5, 8  \\
$\tau^3$ Eri & 18978  & ~A4 V   & 120\phm{XX}  & 4.10 & 0.16 & 37.85 & 1.99 & 12.6 & 2.858 & 0.091 & 8195 & $<$27.93\phm{X} & 1, 5, 7  \\
HR 2265   &  43940 & ~A3 V  &  250\phm{XX} & 5.88 & 0.14  &  16.10 & 1.91 & 13.4 & 2.853 & 0.073 & 8155  & 30.27 & 2, 4, 5  \\
$\iota$ UMa\tablenotemark{b,\,{\rm c}}  & 76644  & ~A7 V & 140\phm{XX} & 3.10 & 0.23 & 68.32 & 2.29 & \phn9.5  & 2.843 & 0.104 & 8060 & 28.45 & 1, 2, 5  \\
$\alpha$ Cep & 203280 & ~A7 IV--V & 283\phm{XX} & 2.44 & 0.22 & 66.84 & 1.57 & 18.2 & 2.807 & 0.127 & 7720 & 27.33 & 8, 9 \\
\enddata
\tablecomments{(a) Wide binary; (b) Close binary; (c) Spectroscopic binary; (d) Eclipsing binary.}
\tablerefs{
(1) Abt \& Morrell 1995; (2) H\"{u}nch et al. 1998; (3) Levato 1972; (4) Royer et al. 2002a; (5) Royer et al 2002b; 
(6) Schmitt 1997; (7) Schmitt et al. 1985; (8) Simon et al. 2002; (9) van Belle et al. 2006; (10) This paper. 
}
\end{deluxetable}

\clearpage
\begin{deluxetable}{lccccc}
\tabletypesize{\scriptsize}
\tablecaption{Journal of {\it FUSE}\, Observations \label{tabobs}}
\tablewidth{0in}
\tablecolumns{6}
\tablehead{
\colhead{ Star~~~ } & \colhead{\it FUSE } & \colhead{ UT Date } & \colhead{No. of}
& \colhead{ Exp. Time$^{a}$ } & \colhead{Night Only$^{a}$}\\[-3pt]
\colhead{  } &  \colhead{ Dataset }  &  \colhead{ (yyyy-mm-dd) }   & \colhead{Subexp.}
& \colhead{(ks)} & \colhead{ (ks)}}
\startdata
HD 129791 &  C0380801000 & 2002-04-24 & 2 & \phn6.3  &  \phn  3.5  \\
HD 159312 &  C0381201000 & 2004-09-11 &  4 &  \phn9.8  &  \phn   5.6 \\
33 Boo  &  C0380301000 & 2002-05-24 & 2 &  \phn6.1 &   \phn 4.2  \\
$\omega$ Aur & C0380901000 & 2002-10-09  & 5 & \phn3.6 &   \phn 2.5 \\
$\iota$ Cen     &  A0410505000  & 2000-07-09   & 11~  & 13.3  &  \phn 8.7 \\

$\delta$ Vel & E0750102000 &  2006-07-12 & 2 &   \phn4.0 &   \phn 3.2 \\ 
$\delta$ Vel & E0750103000 &  2006-07-15 & 4 &   \phn8.3 &   \phn 4.2 \\ 
$\beta$ Leo     &  A0410202000  & 2001-04-17    & 7 & \phn6.8 &  \phn 3.9  \\

$\rho$ Oct & C0380402000 & 2002-08-06 & 3 &  \phn3.9 &     \phn 0.0 \\
$\beta$ Ari     &  A0410101000  & 2001-09-03    & 9 & 10.3  & \phn  2.8 \\
$\delta$ Leo    &  A0410303000  & 2000-12-21   &  9 & \phn7.4  &  \phn 1.8 \\

$\tau^3$ Eri    &  A0410606000  & 2001-08-06   &  4 & 12.6  &  \phn 3.0 \\

HD 43940  & C0380101000 & 2002-11-06 & 2 &  \phn5.6 &   \phn 3.7   \\
$\iota$ UMa     &  A0410405000  & 2001-11-04   &  2 & \phn5.7 &  \phn 1.7  \\
$\alpha$ Cep~~~ &  A0410707000  & 2000-08-12   & 8  & 27.5 &   15.0\\
$\alpha$ Cep~~~    &  A0410708000  & 2000-08-12   & 9  & 25.8  & 12.6 \\
\enddata
\tablenotetext{a}{Cumulative accepted exposure times (good time intervals) at the wavelength of O\,{\sc vi}.}
\end{deluxetable}

\clearpage
\begin{deluxetable}{lcccccc}
\tabletypesize{\scriptsize}
\tablecaption{Observed FUV Emission-Line Fluxes \label{tabfits}}
\tablewidth{0in}
\tablecolumns{7}
\tablehead{
\colhead{ Star~~~ } & \colhead{$f$(O\,{\sc vi})\tablenotemark{a}}& \colhead{$f$(O\,{\sc vi})\tablenotemark{a}}& 
\colhead{$L$(O\,{\sc vi})\tablenotemark{b}} & \colhead{$L$(O\,{\sc vi})/\lbol\tablenotemark{c}} & 
\colhead{$f$(C\,{\sc iii})\tablenotemark{a}}& \colhead{HWHM\tablenotemark{d}/} \\ [-3pt]
\colhead{} & \colhead{1032\,\AA} & \colhead{1038\,\AA} & \colhead{1032\,\AA} & \colhead{1032\,\AA} & 
\colhead{977\,\AA} & \colhead{$v \sin i$}
}
\startdata
HD 129791    & 1.1$\pm$0.3      & \nodata\phn & 22$\pm$6\phn      & 2.0$\pm$0.6   & \nodata\phn  & 0.36 \\
HD 159312    & 1.3$\pm$0.1      & 1.9$\pm$0.1\tablenotemark{e} & 17$\pm$1\phn      & 1.7$\pm$0.1   & \nodata\phn  & \nodata\phn \\
33 Boo       & 1.9$\pm$0.1      & 1.1$\pm$0.1 & 8.3$\pm$0.4   & 0.92$\pm$0.05 & 1.2$\pm$0.2  & 0.50 \\
$\omega$ Aur & 4.7$\pm$0.2      & 2.8$\pm$0.2 & 13$\pm$1\phn  & 1.6$\pm$0.1   & 8.0$\pm$2.0  & 0.27 \\
$\iota$ Cen  & $<0.4$\phn       & $<$0.5\phn  & $<$0.2\phn    & $<$0.02\phn   & \nodata\phn  & \nodata\phn \\
$\delta$ Vel & 7.3$\pm$0.2      & 3.8$\pm$0.6 & 5.2$\pm$0.1   & 0.16$\pm$0.01 & \nodata\phn  & 0.56 \\
$\beta$ Leo  & $<0.6$\phn       & $<$0.4\phn  & $<$0.1\phn    & $<$0.02       & \nodata\phn  & \nodata\phn \\
$\rho$ Oct   & 12.4$\pm$0.6      & 5.6$\pm$0.8 & 66$\pm$3\phn     & 8.3$\pm$0.4   & \nodata\phn  & 0.43 \\
$\beta$ Ari  & 5.2$\pm$0.2      & 2.3$\pm$0.2 & 2.1$\pm$0.1   & 0.22$\pm$0.01 & 7.5$\pm$0.5  & 0.44 \\
$\delta$ Leo & $<0.4$\phn       & $<$0.2\phn  & $<$0.2\phn    & $<$0.02\phn   & \nodata\phn  & \nodata\phn \\
$\tau^3$ Eri & 1.8$\pm$0.1      & 1.3$\pm$0.2 & 1.5$\pm$0.1   & 0.31$\pm$0.02 & \nodata\phn  & 1.40 \\
HD 43940     & 1.7$\pm$0.2      & 1.2$\pm$0.2 & 7.9$\pm$0.9   & 1.5$\pm$0.2   & 2.0$\pm$0.5  & 0.26 \\
$\iota$ UMa  & 3.1$\pm$0.2      & 1.6$\pm$0.1 & 0.80$\pm$0.05 & 0.22$\pm$0.01 & \nodata\phn  & 0.20 \\
$\alpha$ Cep & 10.7$\pm$0.7\phn & 7.3$\pm$1.3 & 2.87$\pm$0.19 & 0.41$\pm$0.03 & 32.5$\pm$2.0\phn & 1.40 \\  [6pt]
\hline\\[-12pt] 
AB Dor\tablenotemark{f} & 44.9$\pm$4.5\phn & $<$25.4\phn      &  11.9$\pm$1.2\phn &  81.5$\pm$8.2\phn & \nodata\phn & \nodata\phn \\
AU Mic\tablenotemark{f} & 20.9$\pm$2.4\phn & 10.6$\pm$1.1\phn & 2.5$\pm$0.3       &  53.3$\pm$6.1\phn & \nodata\phn & \nodata\phn \\
\enddata
\tablenotetext{a}{Emission line flux in units of 10$^{-14}$ \ecs\ at Earth,
not corrected for interstellar extinction.}
\tablenotetext{b}{Emission-line luminosity in units of 10$^{27}$ \es.}
\tablenotetext{c}{Normalized O\,{\sc vi} $\lambda$1032 luminosity in units of 10$^{-7}$.}
\tablenotetext{d}{Ratio of the half-width at half-maximum of O\,{\sc vi} 1032~\AA, 
expressed as a velocity, to the stellar $v \sin i$.}
\tablenotetext{e}{Most of this is likely due to C\,{\sc ii} emission; see text.}
\tablenotetext{f}{\fuse fluxes for AB Dor (K0-1\,IV/V) and AU Mic (M1.6\,Ve) from Redfield et al. 2002.}
\end{deluxetable}


\clearpage
\begin{deluxetable}{lcccc}
\tabletypesize{\scriptsize}
\tablecaption{Comparison of A Stars with Two Active Low-Mass Stars \label{tabtruth}}
\tablewidth{0in}
\tablecolumns{5}
\tablehead{
\colhead{ }  &  \colhead{  }  & \colhead{Known or}  &  \colhead{ }  & \colhead{  }    \\[-3pt]
\colhead{  }   & \colhead{$T_{\rm eff}$}  &  \colhead{Possible} & 
\colhead{\lx(A$_\star$) $\sim$}      &  \colhead{L$_{{\rm O}\,VI}$(A$_\star$) $\sim$} \\[-3pt]
\colhead{Star~~~}  &  \colhead{(K)}  &  \colhead{Binary}   &
\colhead{\lx(dK$_\star$/dM$_\star$)} &  \colhead{L$_{{\rm O}\,VI}$(dK$_\star$/dM$_\star$)}    }
\startdata
HD 129791            &  10,000\phn     & \nodata & $\surd $  & $\surd$  \\
HD 159312            &  9,790               & \nodata & $\surd $  & $\surd$  \\
33 Boo                   &  9,630                & \nodata & $\surd $  & $\surd$  \\
$\omega$ Aur        &  9,570              & $\surd$ & $\surd $  & $\surd$  \\
$\iota$ Cen            & 9.395               & \nodata & \nodata  & \nodata  \\

$\delta$ Vel           & 9,250               & $\surd$ & $\surd $  & $\surd$  \\
$\beta$ Leo           & 8,645               & \nodata & \nodata  & \nodata  \\

$\rho$ Oct            & 8,440                & \nodata & $\surd $  & $\surd$  \\
$\beta$ Ari            & 8,370              & $\surd$ & \nodata  & $\surd$  \\
$\delta$ Leo          & 8,300             & \nodata & \nodata  & \nodata  \\
$\tau^3$ Eri           &  8,195            & \nodata & \nodata   & $\surd$  \\

HD 43940             & 8,155              & \nodata & $\surd $  & $\surd$  \\
$\iota$ UMa          & 8,060             & $\surd$ & \nodata  & \nodata  \\
$\alpha$ Cep~~    & 7,720            & \nodata & \nodata  & $\surd$  \\  
\enddata
\tablecomments{The late-type comparison stars are AB Dor (dK; \teff$\approx5200$ K) and AU Mic (dM; \teff$\approx3500$ K).}
\end{deluxetable}



\end{document}